# Evidence for Majorana bound state in an iron-based superconductor

Dongfei Wang[1,2,†], Lingyuan Kong[1,2,†], Peng Fan[1,2,†], Hui Chen[1], Shiyu Zhu[1,2], Wenyao Liu[1,2], Lu Cao[1,2], Yujie Sun[1,3], Shixuan Du[1,3,4], John Schneeloch[5], Ruidan Zhong[5], Genda Gu[5], Liang Fu[6], Hong Ding[1,2,3,4*], and Hong-Jun Gao[1,2,3,4*]

[1]Beijing National Laboratory for Condensed Matter Physics and Institute of Physics, Chinese Academy of Sciences, Beijing 100190, China

[2]School of Physical Sciences, University of Chinese Academy of Sciences, Beijing 100190, China

[3]CAS Center for Excellence in Topological Quantum Computation, University of Chinese Academy of Sciences, Beijing 100190, China

[4]Collaborative Innovation Center of Quantum Matter, Beijing 100190, China

[5]Condensed Matter Physics and Materials Science Department, Brookhaven National Laboratory, Upton, New York 11973, USA

[6]Department of Physics, Massachusetts Institute of Technology, Cambridge, Massachusetts 02139, USA

†These authors contributed equally to this work
*Correspondence to: dingh@iphy.ac.cn, hjgao@iphy.ac.cn

**The search for Majorana bound state (MBS) has recently emerged as one of the most active research areas in condensed matter physics, fueled by the prospect of using its non-Abelian statistics for robust quantum computation. A highly sought-after platform for MBS is two-dimensional topological superconductors, where MBS is predicted to exist as a zero-energy mode in the core of a vortex. A clear observation of MBS, however, is often hindered by the presence of additional low-lying bound states inside the vortex core. By using scanning tunneling microscope on the newly discovered superconducting Dirac surface state of iron-based superconductor $FeTe_{1-x}Se_x$ ($x$ = 0.45, superconducting transition temperature $T_c$ = 14.5 K), we clearly observe a sharp and non-split zero-bias peak inside a vortex core. Systematic studies of its evolution under different magnetic fields, temperatures, and tunneling barriers strongly suggest that this is the case of tunneling to a nearly pure MBS, separated from non-topological bound states which is moved away from the zero energy due to the high ratio between the superconducting gap and the Fermi energy in this material. This observation offers a new, robust platform for realizing and manipulating MBSs at a relatively high temperature.**

Majorana bound state (MBS) in condensed matter systems has attracted tremendous interest due to its non-Abelian statistics and potential applications in topological quantum computation *(1,2)*. MBS is theoretically predicted to emerge as a spatially localized zero-energy mode in certain *p*-wave topological superconductors in one and two dimensions *(3,4)*. While the material realization of such *p*-wave superconductors has remained elusive, novel platforms for MBS have recently been proposed using heterostructures between conventional *s*-wave superconductors and topological insulators *(5)*, nanowires *(6 - 8)*, quantum anomalous Hall insulator *(9)*, or atomic chains *(10)*, where the proximity effect on a spin-non-degenerate band creates a topological superconducting state. While various experimental signatures of MBS *(11 – 14)*, or Majorana chiral mode *(15)* have been observed in artificial systems, a clear detection and manipulation of MBS are often hindered by the mixing of non-topological bound states and complications of material interface.

Very recently, using high-resolution angle-resolved photoemission spectroscopy (ARPES), we have discovered a potentially new platform for MBS in a bulk superconductor FeTe$_{0.55}$Se$_{0.45}$ ($T_c$ = 14.5 K) with a simple crystal structure (Fig. 1A). Due to the topological band inversion between the $p_z$ and $d_{xz}/d_{yz}$ bands around the Γ point *(16, 17)* and the multi-band nature (Fig. 1B), this single material naturally has a spin-helical Dirac surface state with an induced full superconducting (SC) gap and a small Fermi energy *(18)* (Fig. 1C), which are favorable conditions for observing a pure MBS *(5)*, isolated from other non-topological Caroli-de Gennes-Matricon states (CBSs) *(19,20)*. The combination of high-$T_c$ superconductivity and Dirac surface state in a single material removes the challenging interface problems in previous proposals, and offers clear advantages for detecting and manipulating MBS.

Motivated by the above considerations, we carry out a high-resolution scanning tunneling microscopy/spectroscopy (STM/S) experiment on the surface of FeTe$_{0.55}$Se$_{0.45}$ ($T_c$ = 14.5 K), which has a good atomic resolution revealing the lattice formed by Te/Se atoms on the surface (Fig. 1D). We start with a relatively low magnetic field of 0.5 T along the *c*-axis at a low temperature of 0.55 K, with a clear observation of vortex cores in Fig. 1E. At the vortex center, we observe a strong zero-bias peak (ZBP) with a full width at half maximum (FWHM) of 0.3 meV and an amplitude of 2 relative to the intensity just outside the gapped region. Outside of the vortex core, we clearly observe a superconducting spectrum with multiple gap features, similar to the ones observed by previous STM studies on the same material *(21,22)*. These different SC gaps correspond well with the SC gaps on different Fermi surfaces of this material observed by previous ARPES studies *(23)* (more details in table I of Supplementary Information (SI)). We note a similar ZBP was reported previously *(22)*.

We next demonstrate in Fig. 2 and Fig. S4 that under a large range of magnetic field the observed ZBP does not split when moving away from a vortex center. We carry out *dI/dV* measurements along the line going through the bright spot inside the vortex core (Fig. 2A), and display a *dI/dV* intensity plot, a *dI/dV* waterfall-like plot, and a selective *dI/dV* overlapping plot in Figs. 2B – 2D, respectively. It is visually evident that the ZBP remains at the zero energy while its intensity fades away when moving away from the vortex center. The non-split ZBP contrasts sharply with the split ZBP originating from CBS observed in conventional superconductors *(19, 20)*, and is consistent with tunneling into an isolated Majorana bound state in a vortex core of a topological superconductor *(5, 24 - 26)*.

We then extract the values of ZBP height and width from a simple Gaussian fit (Fig. 2E), demonstrating a decaying spatial profile with a nearly constant linewidth of about 0.3 meV in the middle, close to the total width (~ 0.28 meV) contributed from the STM energy resolution (~ 0.23 meV as shown in SI) and the thermal broadening ($3.5k_BT$ @ 0.55 K ~ 0.17 meV). We further compare the observed ZBP height with a theoretical MBS spatial profile obtained by solving the Bogoliubov-de Gennes equation analytically *(5,24)* or numerically *(25,26)*. By using the parameters of $E_F$ = 4.4 meV, $\Delta_{sc}$ = 1.8 meV, and $\xi_0 = v_F/\Delta_{sc}$ = 12 nm, which are obtained directly from the topological surface state by our STS and ARPES results *(18)* (Fig. 2F), the theoretical MBS profile matches remarkably well with the experimental one (Fig. 2G).

The observation of a non-split ZBP, which is different from the split ZBP observed in a vortex of the $Bi_2Te_3/NbSe_2$ heterostructrue, indicates that the MBS peak in our system is much less contaminated by other non-topological CBS peaks, which is made possible by the large $\Delta_{sc}/E_F$ ratio. In a usual topological insulator/superconductor heterostructure, this ratio is tiny, on the order of $10^{-3} - 10^{-2}$ *(27)*. This has been shown to induce, in addition to the MBS at the zero energy, many CBSs, whose level spacing is proportional to $\Delta_{sc}^2/E_F$. As a result, these CBSs crowdedly pack very close to the zero energy, obscuring a clean detection of MBS from the *dI/dV* spectra *(28)*. However, on the surface of $FeTe_{0.55}Se_{0.45}$, the $\Delta_{sc}/E_F$ ratio is about 0.4, which is sufficiently large to push most CBSs away from the zero energy (more details are in SI), leaving the MBS largely isolated and unspoiled. We also note that all the bulk bands in this multi-band material have fairly small values of $E_F$ due to large correlation-induced mass renormalization, ranging from a few to a few tens meV, thus their values of $\Delta_{sc}^2/E_F$ are also quite large (> 0.2 meV as shown Table S1 of SI). These large bulk ratios enlarge the energy level spacing of CBSs inside the bulk vortex line, which helps reducing quasiparticle poisoning of the MBS at low temperature (more details are in SI).

It has been predicted *(29)* that the width of the zero-bias peak from tunneling into a single isolated MBS is determined by thermal smearing ($3.5k_BT$) and tunneling broadening. Additionally, STS has its own energy resolution due to instrumentation limitation. We measure tunneling barriers evolution of the ZBP (Fig. 3A). Robust ZBPs can be observed over two orders of magnitude in tunneling barrier conductance, with the width barely changed (Fig. 3B). We also note the linewidth of ZBPs is almost completely limited by the combined broadening of energy resolution and STM thermal effect, indicating that the intrinsic width of the MBS is much smaller, and our measurements are within the weak tunneling regime.

However, we do observe some other ZBPs with larger broadening (Fig. 3C). Interestingly, a larger ZBP broadening is usually accompanied with a softer superconducting gap. We summarize this correlation in Fig. 3C: the FWHM of ZBP increases with increasing sub-gap background conductance measured at the edge of vortex. The sub-gap background conductance, which may vary with different scattering strength from disorders, quasiparticle interactions and others *(30 - 32),* introduces a gapless fermion bath that can poison the MBS, as explained previously *(33)*. The effect of quasiparticle poisoning is to reduce the MBS amplitude and increase its width. This scenario is likely the origin of a larger broadening of ZBP accompanied by a softer gap.

It has been pointed out by previous theoretical studies *(34 - 36)* that the condition of a bulk vortex line has significant influences to the Majorana mode on the surface. In order to further characterize the influences of bulk vortex line, we have performed the temperature evolution of a ZBP. As shown in Fig. 3D, the ZBP intensity measured at a vortex center decreases with increasing temperature, and becomes invisible at 4.2 K. If it would be associated with a CBS, it should last at higher temperatures and may exhibit simple Fermi-Dirac broadening up to about $T_c/2$ (about 8 K) below which the superconducting gap amplitude is almost constant, as observed in our previous ARPES measurement *(18)*. Our observation (Fig. S5) contradicts this expectation and indicates an additional suppression mechanism that is likely related to poisoning MBS by thermally excited quasiparticles. From the extraction of ZBPs amplitude measured on three different vortices (Fig. 3E), we find that the observed ZBPs vanish around 3 K, which is much higher than the temperature in many previous Majorana platforms *(11, 37)*. This vanishing temperature is comparable to the energy level spacing of the bulk vortex line as discussed above. This is consistent with a case of a MBS poisoned by thermal-induced quasiparticles inside the bulk vortex line (more details are in SI). Under the condition of sufficiently low temperature and free of other Majorana poisoning sources, the isolated MBSs in this system can be used to demonstrate non-Abelian braiding (Fig. 3F).

To summarize our findings: 1) we have observed a non-split ZBP inside a vortex core of $FeTe_{0.55}Se_{0.45}$, which is robust under a magnetic field up to 6 T; 2) the spatial profile of ZBP is consistent with theoretical predictions; 3) at low temperature the linewidth of ZBP is resolution and temperature limited, and remains a constant over two orders of magnitude change in tunneling barrier, at high temperature there is an additional peak suppression mechanism that can be attributed to quasiparticle poisoning. Our observations demonstrate strong evidence that this is a case of tunneling to an isolated Majorana bound state, and many other trivial explanations, as discussed in SI, cannot account for all the observed features. The high transition temperature and large superconducting gaps in this superconductor bring a new and widely available venue for Majorana research based on natural-occurring quantum materials, and offer a promising platform to fabricate robust devices for topological quantum computation.

**Acknowledgements:**
We thank Qing Huan, Hiroki Isobe, Xiao Lin, Xu Wu, Kai Yang for technical assistance, and Patrick A. Lee, Tai Kai Ng, Shuheng Pan, Gang Xu, Jia-Xin Yin, Fuchun Zhang, Peng Zhang for useful discussions. This work at IOP is supported by grants from the Ministry of Science and Technology of China (2013CBA01600, 2015CB921000, 2015CB921300, 2016YFA0202300), the National Natural Science Foundation of China (11234014, 11574371, 61390501), and the Chinese Academy of Sciences (XDPB08-1, XDB07000000, XDPB0601). L.F. is supported by the U.S. Department of Energy under Award DE-SC0010526. G.D.G is supported by the U.S. Department of Energy under Contract No. DE-SC0012704. J.S. and R.D.Z. are supported by the Center for Emergent Superconductivity, an Energy Frontier Research Center funded by the U.S. Department of Energy.

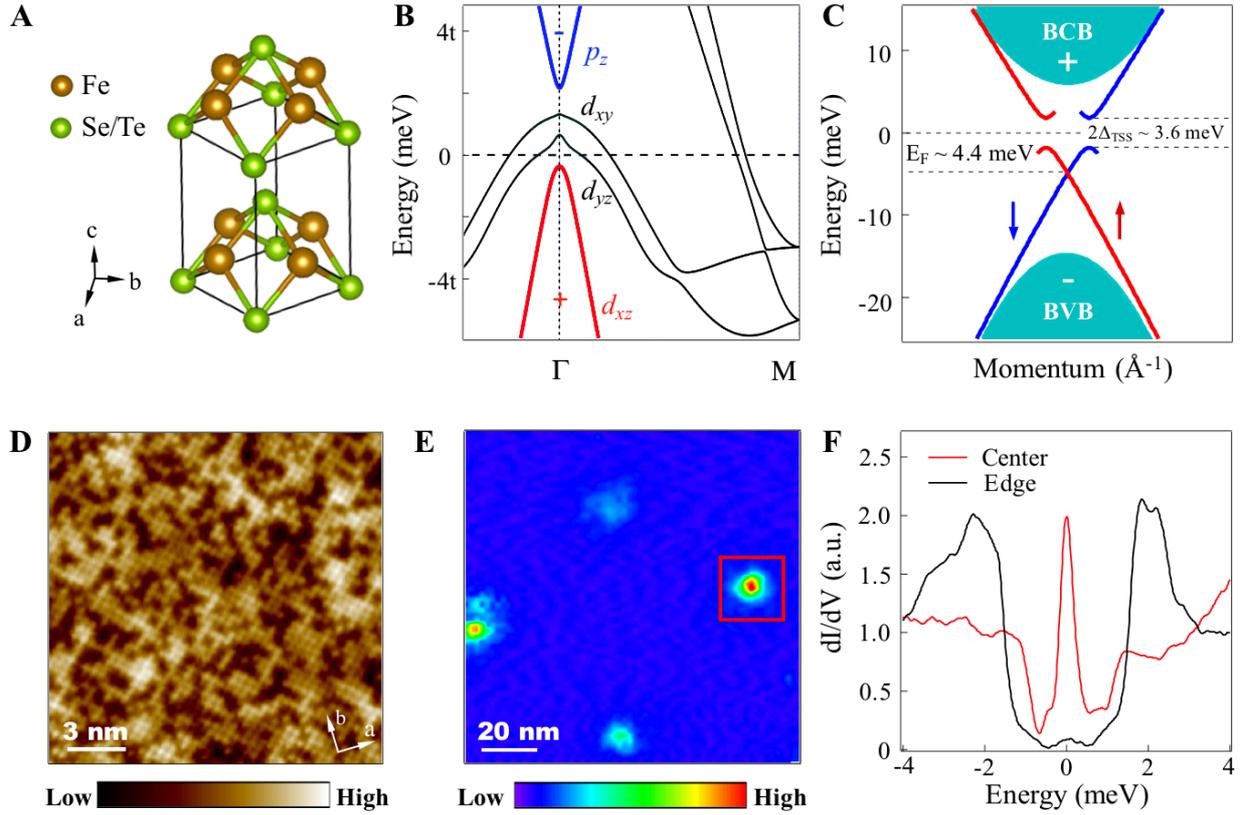

Fig. 1 **Band structure and vortex cores of FeTe$_{0.55}$Se$_{0.45}$.** (A) Crystal structure of FeTe$_{0.55}$Se$_{0.45}$. Axis **a** or **b** indicates one of Fe-Fe bond directions. (B) *ab initio* calculation of the band structure along the Γ-M direction, adopted from Fig. 1C of Ref *(18)*. It shows un-inverted bands structure at $k_z = 0$ plane. In the calculations, $t = 100$ meV, while $t \sim 12 - 25$ meV from ARPES experiments, largely depending on the bands *(23)*. (C) Summary of superconducting topological surface states on this material observed by ARPES from Ref. *(18)*. (D) STM topography of FeTe$_{0.55}$Se$_{0.45}$ (scanning area: 17 nm × 17 nm). (E) Normalized zero-bias conductance (ZBC) map measured at a magnetic field of 0.5 T, with the area 120 nm × 120 nm. (F) A sharp ZBP in a *dI/dV* spectrum measured at the vortex core center indicated in the red box on (E). It is in sharp contrast to the spectrum measured at the edge of the vortex where a two-gap feature is largely unchanged from the case of zero field. Settings: sample bias, $V_s = -5$ mV; tunneling current, $I_t = 200$ pA; magnetic field, $B_\perp = 0.5$ T; and temperature, $T = 0.55$ K.

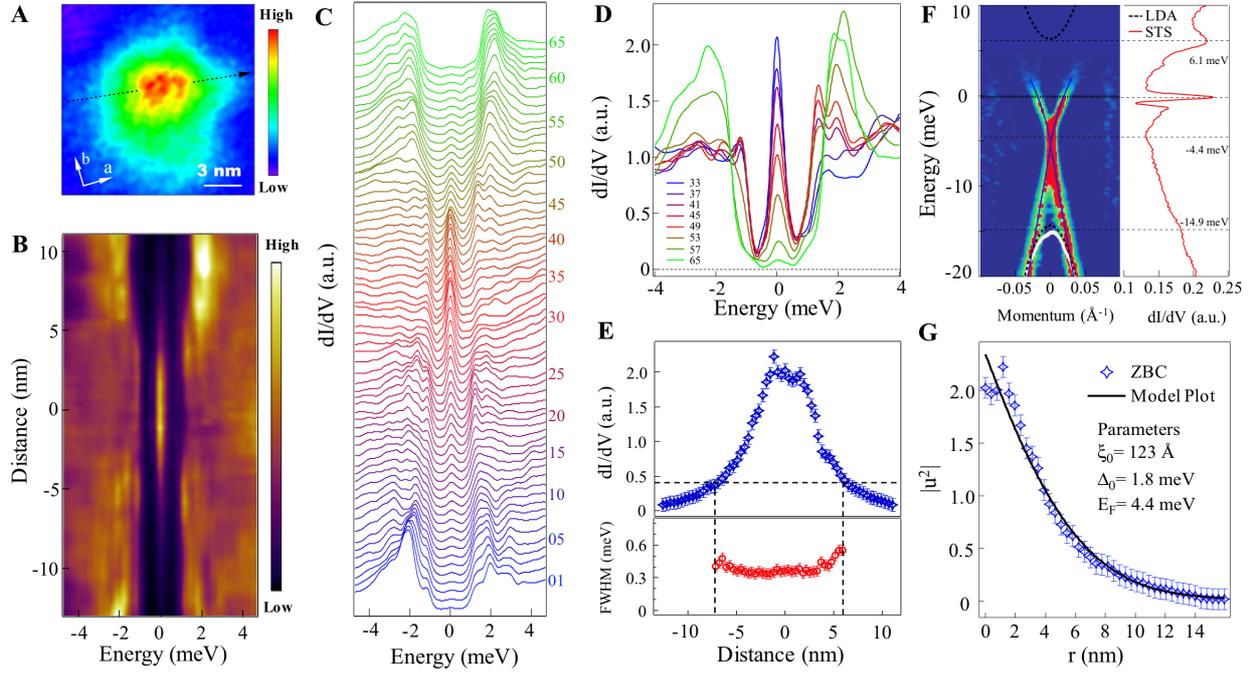

Fig. 2 **Energetic and spatial profile of Majorana bound states.** (**A**) A high resolution ZBC map (area = 15 nm × 15 nm) around vortex cores. (**B**) A line-cut intensity plot along the black dash line indicated in (A). (**C**) A waterfall-like plot of (B) with 65 spectra. (**D**) An overlapping display of 8 selected *dI/dV* spectra. (**E**) Spatial profile (upper panel) and FWHM (lower panel) of the ZBP. (**F**) Comparison between ARPES and STS results. Left panel: ARPES results on the topological surface states adopted from Ref. *(18)*. Black dash curves are extracted from a LDA calculation *(36)*, with the LDA data rescaled to match the energy positions of the Dirac point and the top of bulk valance band (BVB). Right panel: a *dI/dV* spectrum measured from -20 meV to 10 meV. Note that the bump at -14.9 meV in STS corresponds to the top of bulk valence band in ARPES, the dip at -4.4 meV in STS matches to the Dirac crossing in ARPES, and a peak at + 6.1 meV in STS matches to the bottom of the bulk conduction band. (**G**) Comparison between the measured ZBP peak intensity with a theoretical calculation of MBS spatial profile. Data in (B)-(G) is normalized by integrated area of each *dI/dV* spectra. Settings: $V_s$ = -5 mV, $I_t$ = 200 pA, and $T$ = 0.55 K, $B_\perp$ = 0.5 T.

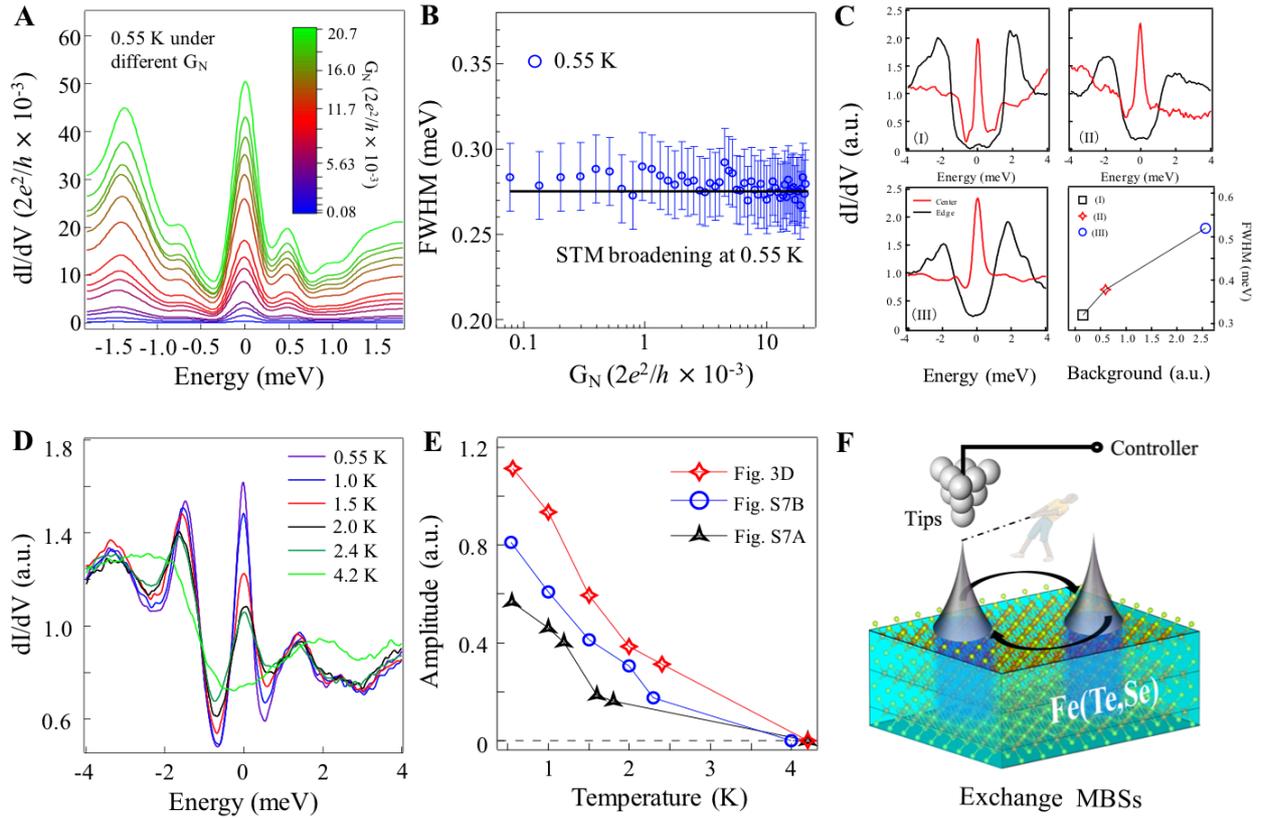

Fig. 3 **Temperature and tunneling barrier evolution of Majorana bound states.** **(A)** Tunneling barrier evolution of ZBPs measured at 0.55 K. $G_N \equiv I_t/V_s$, which corresponds to the energy-averaged conductance of normal states, and represents conductance of tunneling barrier. $I_t$ and $V_s$ are the STS setpoint parameters. **(B)** FWHM of ZBPs at 0.55 K under different tunneling barriers. The black solid line is the combined effect of energy resolution (0.23 meV as shown in SI) and tip thermal broadening ($3.5k_BT$) at 0.55 K. **(C)** FWHM of ZBP at the center of vortex core is larger when the superconducting gap around the vortex core is softer. Background is defined as an integrated area from -1 meV to +1 meV of the spectra at the core edge. **(D)** Temperature evolution of ZBPs in a vortex core. **(E)** Amplitude of the ZBPs shown in (D) and Fig. S7 of SI under different temperatures. The amplitude is defined as the peak-valley difference of the ZBP. **(F)** Schematic of a possible way for realizing non-Abelian statistics in a future ultra-low-temperature STM experiment which may have an ability to exchange MBSs on the surface of Fe(Te, Se). Settings: (A) and (B) are in absolute value of conductance, $B_\perp$ = 2.5 T. (D) and (E) are normalized by integrated area, $V_s$ = -10 mV, $I_t$ = 100 pA, $T$ = 0.55 K, $B_\perp$ = 4 T.

## Supplementary Information for

### Evidence for Majorana bound state in an iron-based superconductor

Dongfei Wang[†], Lingyuan Kong[†], Peng Fan[†], Hui Chen, Shiyu Zhu, Wenyao Liu, Lu Cao, Yujie Sun, Shixuan Du, John Schneeloch, Ruidan Zhong, Genda Gu, Liang Fu, Hong Ding[*], and Hong-Jun Gao[*]

[†]These authors contributed equally to this work
[*]Correspondence to: dingh@iphy.ac.cn, hjgao@iphy.ac.cn

**This PDF file includes:**

    Materials and Methods

    Supplementary Text
- I. Voltage offset and energy resolution calibrations
- II. Topological nontrivial and strong correlated properties of bulk bands
- III. Discussions on several trivial explanations of observed zero-bias peak
- IV. Intrinsic inhomogeneity and different types of bound state inside vortex core
- V. Non-split zero-bias peak under different magnetic field
- VI. Additional results of temperature evolution
- VII. Additional results of tunneling barrier evolution
- VIII. Details of MBS wave function calculation.

    Figs. S1 to S8
    Tables S1

**Materials and Methods:**

Large single crystals of FeTe$_{0.55}$Se$_{0.45}$ with high quality were grown using the self-flux method, and their values of $T_c$ were determined to be 14.5 K from magnetization measurements *(1)*. There are two kinds of single crystals crystallizing simultaneously with similar structure and Te/Se compositions. Fe$_{1+y}$Te$_{0.55}$Se$_{0.45}$ single crystals with excess Fe atoms, with shinning surfaces and being easy to cleave, are non-superconducting before annealing under Te atmosphere. FeTe$_{0.55}$Se$_{0.45}$ single crystals without excess Fe, usually without shinning surface, are superconducting without post-annealing. All STM/STS data shown in this paper are from as-grown FeTe$_{0.55}$Se$_{0.45}$ single crystals, and previous ARPES data which shows topological surface state are also from this kind of samples *(2)*. The samples used in the experiments were cleaved *in situ* and immediately transferred to a STM head. Experiments were performed in two different ultrahigh vacuum (1 × 10$^{-11}$ mbar) LT-STM systems, STM#1 (USM-1300s-$^3$He) and STM#2 (USM-1300-$^3$He with a vector magnet), STM images were acquired in the constant-current mode with a tungsten tip. Differential conductance (*dI/dV*) spectra were acquired by a standard lock-in amplifier at a frequency of 973.1 Hz, under modulation voltage $V_{mod}$ = 0.1 mV. The tips were calibrated on a clean Nb(111) surface prepared by repeated cycles of sputtering with argon ions and annealing at 1200 $^\circ$C. Low temperatures of 0.4 K (0 T) and 0.55 K (4 T) are achieved by a single-shot $^3$He cryostat. A perpendicular magnetic field up to 11 Tesla for STM#1 and a vector magnetic field with the maximum value 9$_z$-2$_x$-2$_y$ Tesla for STM#2 can be applied to a sample surface. Data in part (II) & (III) of Fig. 3C, Figs. 3D and E, Figs. S2A-C and F-H, Figs. S3A-B and E-F, Figs. S4F and H (middle panels), Figs. S5-S7 were measured by STM#1, while others were from STM#2.

**I. Voltage offset and energy resolution calibrations**

It is well known that the zero-bias offset problem exists in STM studies, and it is critically important in our study of the zero-energy peaks of MBS. We followed a standard procedure *(3)* in the field with extra cares. We calibrated the two STM systems with the standard method of overlap spot of *I-V* curves, since the current should be always zero when the voltage is zero. An example of this calibration is shown in Figs. S1 (A)-(B), which shows the zero-bias offset of 57.7 μeV. This offset was then subtracted from all the following spectra under the same condition. We have recalibrated the STM several times during a cycle of measurement (usually in 17 hours) and found no detectable drift during the cycle. We have further confirmed this calibration by checking the particle-hole symmetry of the superconducting gap, as shown in Figs. S1 (C)-(D).

We calibrated the energy resolution by measuring a superconducting Nb single crystal. As shown in Fig. S1 (C), the overall broadening of the superconducting spectrum edge is about 0.3 meV for STM#1 and 0.26 meV for STM#2. By removing the temperature broadening term (0.12 meV for 0.4 K without applying a magnetic field), it shows that the energy resolution is better than 0.27 meV for STM#1 and 0.23 meV for STM#2 *(4)*. The total broadening at 0.55 K is about 0.32 meV for STM#1 and 0.28 meV for STM#2.

$$Total\ broadening = \sqrt{r_{Temperature}^2 + r_{System}^2 + r_{Scattering}^2 + r_{Others}^2} \qquad (Eq.\ S1)$$

## II. Topological nontrivial and strong correlated properties of bulk bands

Fe(Te, Se) is a typical '11-type' iron-based superconductor with the simplest crystal structure among all the iron-based superconductors. It has three hole-like bands around the Γ point and two electron-like bands around the M point in the bulk bands *(5)*. Topological nontrivial band structure occurs around the Γ−Z high symmetric line when the $p_z/d_{xy}$ anti-bonding orbital with an odd parity inverts with the $d_{xz}/d_{yz}$ orbital which is even parity. This band inversion can be controlled by Te atom substitution. With increase of Te atoms, the interlayer *p-p* orbital hybridization is strengthened while the intra-layer $p_z$-$d_{xy}$ hybridization is weakened. In addition, Te substitution also increases spin orbital coupling (SOC), which induces a gap opening at the band cross. Through this process, Fe(Te, Se) gains a nontrivial $Z_2$ topological invariance, whose surface would have a topological surface state that has a helical spin structure *(6)*, which has been clearly observed by a recent high-resolution spin-resolved ARPES experiment *(2)*.

Fe(Te, Se) is also known to possess the strongest electron correlations among all the iron-based superconductors, with the mass renormalization of ~ 8 for $d_{xy}$ orbital and ~ 5 for $d_{xz}/d_{yz}$ orbitals *(7-9)*. This heavily reduces the bandwidths and consequently the Fermi energies for all the bulk bands. An important consequence of small bulk Fermi energies is that the energy level spacing of Carlo-de Gennes-Matricon states inside the bulk vortex line, which is $\Delta^2/E_F$, is large enough (about 0.2 meV or larger, see Table S1) to allow a clear detection of MBSs at the low temperature (0.55 K) used in our experiment.

## III. Discussions on several trivial explanations of the observed zero-bias peak

A zero-bias conductance peak robust against perturbations is a necessary piece of evidence for a Majorana bound state in materials, which gives rise to resonant Andreev tunneling at the zero energy *(10)*. Following theoretical proposals, zero-bias conductance peak has recently been observed in semiconductor nanowire devices *(11-15)* and ferromagnetic atomic chains *(16-18)*. However, the energy gap separating the zero-energy MBS and excited quasiparticle states, which is required for the topological protection of MBS, is yet to be clearly demonstrated in both systems. Moreover, spatial profile of MBS in nanowires has not been measured. In the vortex version, a 2D surface provides a good platform not only to measure the pronounced zero-bias peak, but also to check detailed the spatial structure of Majorana wave function, that is, the spatial profile of zero-bias peak *(19,20)*.

It has been pointed out that several effects may cause a zero-bias peak in tunneling experiments, such as weak antilocalization *(21, 22)*, reflectionless tunneling *(23)*, Kondo effect *(24, 25)*, Josephson supercurrent *(26, 27)* and packed CBS states near the zero energy *(28-31)*. Many of these alternative effects, such as reflectionless tunneling, involve Andreev reflection process between a normal lead and a disordered superconductor *(32)*. However, Andreev process is significantly suppressed at the center of a vortex where the amplitude of the superconducting is forced to vanish *(33)*. Other trivial explanations, as discussed in details below, are contradicted with our experiment observations.

1) *Reflectionless tunneling* is due to electron/hole phase conjugation through multiple Andreev reflection process, in which scattering centers make mirror-reflected electrons shoot back to the sample again. This effect can be ruled out by a relatively high magnetic field used in our STM experiments as the extra phase accumulation induced by the magnetic field would destroy the phase conjugation condition. In the meanwhile, the vacuum tunneling configuration of STM does not benefit such a re-incident process, and the suppression of Andreev process on the vortex center also frustrates this effect.

2) *Kondo effect* can be ruled out by the zero field measurement shown in Figs. S4E & S4G, since there are no visible impurities in the areas selected for vortex measurement. Instead, our measurement shows a hard superconducting gap. In addition, sharp ZBPs appear over a large range of magnetic field as shown in Figs. S4F & S4H, which contradicts with the splitting of Kondo resonance peaks at high magnetic field *(34)*.

3) *Josephson supercurrent* will manifest itself as a zero-bias peak in a conductance spectrum under a small tunneling barrier. However, the tungsten tip used in our experiment retains metallic at low temperature, and without containment from the superconducting sample, so there is no SIS junction that is required in Josephson tunneling. In addition, the zero-bias peak can be detected under a large tunneling barrier, which further rules out this explanation.

4) *Weak anti-localization (localization) effect* can cause a zero-bias conductance peak (dip) in the quantum diffusive transport region, where a phase destructive (constructive) interference of backscattered electrons is protected by the time reversal symmetry. Conventional weak anti-localization (WAL) was widely studied in topological insulators *(35)*. This effect can be ruled out by the high magnetic field used in our experiment. But the Andreev process could provide a different mechanism for phase conjugation, no matter the time reversal symmetry holds or not. WAL can coexist with a magnetic field when tunneling from a normal lead to a superconductor. However, as we mentioned above, the condition for this effect is not satisfied in our measurements since the pairing strength must be zero at the vortex center and the Andreev process is strongly suppressed *(36)*.

5) *Near-zero-energy packed CBS states* are common in vortices of conventional superconductors. However, as discussed in the main text and in this SI below, this scenario is inconsistent with our observations. First, CBS states inside a vortex core would shift toward the gap edge when moving from the core center to the edge of a vortex. They usually show dispersing features in a spatial line-cut plot as shown in Fig. S3C. It is not consistent with the non-split nature of ZBP we observed. Second, the energy of the lowest CBS is determined by the ratio of $\Delta^2/E_F$, which in our work is considerably large (about 0.7 meV for the surface band and greater than 0.2 meV for the bulk bands). That pushes the CBS away from the zero energy, exactly as we observed in Fig. S3C. Moreover, FWHM of the ZBP observed at low temperature is very sharp, thus the ZBP is unlikely an envelope peak formed by many CBS peaks near the zero energy.

We thus conclude that the above alternative effects cannot explain the ZBP observed in our measurements. Therefore, the most likely mechanism of the ZBP is Majorana quasiparticles enhanced resonant Andreev reflection *(10)*, as one takes a comprehensive consideration regarding its spatial non-split line shape and its behaviors under different temperatures, magnetic fields, and tunneling barriers.

## IV. Intrinsic inhomogeneity and different types of bound state inside vortex core

Figure S3 lists three different cases of bound states we observed inside vortex cores: 1) In some areas with no visible Fe or other impurities, strong MBS are observed inside vortex cores, CBS states is either absent or weak in this case, as shown in Fig. S2 and Figs. S3A & B. 2) There are also cases of coexistence of MBS and CBS, and CBS stays slightly away (usually around 0.2 - 0.7 meV) from the Fermi level at the center of the vortex core and disperses to the higher energy when moving away from the core center, while MBS always stays at the zero energy as shown in Figs. S3 C, D and Fig. S4F. This is consistent with the fact that the energy level spacing of CBS is controlled by $\Delta^2/E_F$, which has a value around 0.7 meV for the topological surface state and a minimal value of 0.2 meV for the bulk bands. In fact, the non-zero peak observed at the core center in Figs. S3C, D is at about 0.7 meV on the empty side, almost identical with the lowest CBS energy calculated with the extracted parameters of the Dirac surface bands in the main text ($\Delta^2/E_F$ = 0.74 meV). The fact that this CBS mode appears in the empty side also supports the origin of the electron-like surface Fermi surface since it is known that the lowest CBS peak appears only on empty (filled) state if the carriers are electron (hole) *(37 - 39)*. 3) In some areas, only CBS can be found inside a vortex core, as shown in Figs. S3 E, F.

Statistically, we have about 20% success rate in observing isolated pure MBS during our more than 150 measurements, and our experience indicates the importance of high quality of single crystals and spatial homogeneity of superconducting gaps. While the exact reason of these variations is not fully understood yet, we offer two possible explanations below.

First, the bulk vortex line has low-lying sub-gap states. In the presence of disorder, the spectral gap inside the vortex line may close. The coupling of MBS to the zero-energy sub-gap states on the vortex line makes it move deeper beneath the surface and consequently reduces the tunneling conductance in STS, which is highly surface sensitive. Nonetheless, provided that the zero-energy states along the vortex line are localized, the MBS at the two ends remain spatially separated, and topologically protected. It may be also possible that disorder is sufficiently strong that drive a topological phase transition for the entire vortex line as suggested by Ref. *(40)*, leading to ZBPs in some of the vortex cores while absent in others.

Second, since Te substitution plays a critical role on the topological nature of Fe(Te, Se), the inhomogeneous distribution of Te/Se, as observed by a previous TEM report *(41)* and our STM (see Fig. 1D), may introduce some non-topological regions in the bulk, thus causing disappearance of topological surface state and consequently MBS on some regions on the surface. In addition, the inhomogeneities of Te/Se substitution and charge doping due to charge transfer effect between chalcogen atoms and iron atoms *(41)* may also modify band dispersions and Fermi energies of both the bulk and surface states, which may affect the condition for detecting a Majorana mode in tunneling spectroscopy.

## V. Non-split zero-bias peak under different magnetic field

Our field-dependent STM experiments reveal that the ZBP starts to emerge at 0.15 T, maintaining its zero-energy position up to 6.0 T, as shown in Fig. S4. With increasing the magnetic field, the inter-vortex length decreases, with the values of 70 nm, 32 nm, 25 nm and 20 nm at 0.5 T, 2.5 T, 4.0 T and 6.0 T, respectively. The amplitude and width of MBSs are quite stable from 0.15 T to 6.0 T, indicating the ZBP in Fe(Te, Se) is very robust against the magnetic field. The relevant small coherence length of MBS shown in the main text enables MBS to be stable at sufficiently high magnetic field. The presence of ZBP and its non-split line profile exclude the possibilities that ZBPs made by Kondo resonance. During our experiment, we carefully checked the selected area by a zero-field mapping before detailed vortex study. We make sure that there is no visible impurity state involved in the signal of zero bias peaks.

**VI. Additional results of temperature evolution**

As shown in Fig. S5, we replot the temperature dependent spectra in Fig. 3D and numerical broadened lowest temperature spectra by convolution with relevant Fermi-Dirac function, we found the absence of ZBPs at 4.2 K is not a simple consequence of thermal broadening.

Typical ZBC maps under different temperatures are shown in Figs. S6A-S6D. At low temperature, vortices are more regular and show a strong conductance on the center. A low temperature linecut intensity plot and a corresponding waterfall plot are shown in Figs. S6E and S6F, respectively. It is similar to the linecuts shown in the main text and Fig. S2 that a spatial non-split pronounced ZBP dominates in a vortex core. At higher temperatures, the vortex becomes more irregular and diffusive. In addition, the conductance intensity at the vortex core decreases rapidly at high temperature. A high temperature linecut intensity plot and a corresponding waterfall plot for the same vortex core are shown in Figs. S6G and S6H, respectively, showing disappearance of a ZBP.

Beyond what has been shown in Fig. 3 of the main text, two more cases of temperature evolution of MBS are shown in Figs. S7A and B, showing similar behaviors as the data in Fig. 3D. By measuring the normalized differential conductance differences between peak and valley, we summarize their temperature dependence for the two cases in Fig. S7C. It can be derived from Eq. S2 in the next part of SI that the conductance difference can be fitted by $C/T$. C is the fitting parameter and represent a characteristic temperature indicating the robustness of ZBP against thermal influence. Such the fits are shown in the left panel of Fig. S7C. As shown in right panel of Fig. S7C, it is evident that C is positively correlated with the low temperature amplitude of ZBP.

This temperature behavior of MBS can be understood in the following way, as schematically illustrated in Fig. S7D, at sufficient low temperature when the quasiparticle thermal energy is smaller than the energy level spacing of CBS inside the bulk vortex line, MBS at the end of vortex line can survive poisoning from thermal excited quasiparticles, as illustrated in the left panel of Fig. S7D. However, at higher temperature when the thermal energy is larger than the level spacing, MBS will be strongly poisoned by the thermally excited quasiparticles, as illustrated in in the upper right panel of Fig. S7D. A simple quantitative analysis can estimate the value of MBS vanishing temperature. As shown in Table S1, the minimal value of the energy level spacing of CBS that is closely related to $\Delta^2/E_F$ of the bulk which has a typical value of about 0.3 meV, compatible to the thermal energy ($k_B T$) around 3K. As shown in Fig. S7C, the extracted C parameter is different from vortex to vortex. It can be understood in the same scenario. As we mentioned in part IV of SI, the intrinsic spatial inhomogeneity of Fe(Te, Se) may allow different types of vortex line exist at different positions.

## VII. Additional results of tunneling barrier evolution

Tunneling barrier dependence data are measured within a fixed 'setpoint' voltage ($V_s$ = -5 mV). Different tunneling barriers are achieved by changing the tunneling current ($I_t$), corresponding to changing of tip/sample spacing, while STM regulation loop is active. As mentioned in the main text, the tunneling barrier is parameterized by the conductance calculated by the STM 'setpoint' settings, that $G_N \equiv I_t/V_s$. $G_N$ is an averaged normal state conductance which represents the conductance of tunneling barrier *(42)*, which is more accurate than the high-bias conductance used in some studies on nanowire devices *(14)*.

Majorana bound states can be observed in tunneling spectrum by resonant Andreev reflection coupled to a normal lead. It intrinsically leads to $2e^2/h$ quantized conductance plateau at zero temperature. Under the conditions of finite temperature and large tunneling barrier, it is difficult to realize $2e^2/h$ by STM configuration. A theoretical model based on resonant tunneling to a single Majorana mode *(44)* shows at low temperature ($k_BT \ll \Delta_0$) and weak tunneling ($\Gamma \ll \Delta_0$) conditions, the zero-bias conductance of a MBS can be scaled onto a simple function (Eq. S2), which only depends on a dimensionless ratio $k_BT/\Gamma$ *(14, 43)*, as shown in the inset of Fig. S8A. $\Gamma$ is the parameter of tunneling broadening, which is the HWHM of a theoretical MBS at 0 K.

$$G_s = \frac{2e^2}{h} \int_{-\infty}^{+\infty} dE \frac{\Gamma^2}{E^2 + \Gamma^2} \frac{1}{4k_BT \cosh^2(E/2k_BT)}$$

$$= \frac{2e^2}{h} f(k_BT/\Gamma)$$

(Eq.S2)

This scaling function gives the value of MBS conductance at a finite temperature with different tunneling barriers. That creates possibilities for us to verify the Majorana nature of the observed zero-bias peak below the quantized conductance region. We tried this strategy in Fig. S8A, where 49 zero-bias conductance of the ZBPs measured at 0.55 K are scaled and compared to the theoretical curve. We note that the relation $\Gamma = \kappa G_N/(2e^2/h)$ holds at the sufficiently weak tunneling condition *(43)*, and we choose the value $\kappa = 170$ as a single adjustable parameter in the scaling process. We find the tunneling broadening in our experiment $G_N$ interval is 2~4 orders of magnitude smaller than the FWHM of MBS (Fig. S8B). It is the reason why the MBS width barely changes over two orders of magnitude in $G_N$. Even the experimental data shows the scaling behavior consistent very well with the theoretical curve, we cannot regard this result as a supporting evidence of Majorana bound states. All of data point shown in Fig. S8A are measured in the weak coupling region in which the linear scaling behavior of STM itself can also lead to this good consistence. It will be more convincing to try this strategy under the condition of lower temperature and lower tunneling barrier (small $k_BT/\Gamma$), which push the experiment beyond the linear region of scaling function.

Tunneling barrier dependence measurement is repeated on another vortex (Figs. S8C & D), in which we obtain a similar behavior.

## VIII. Details of MBS wave function calculation

Details of MBS wave function calculations are as follows. From previous theoretical work *(19,20)*, the intensity of Majorana wave function can be written as

$$|u|^2 = |f(r)|^2 + |g(r)|^2 \quad \text{(Eq.S3)}$$

Where

$$f(r) = J_0(\tfrac{E_F r}{v_F}) \exp\left[-\int^r \tfrac{\Delta(r')}{v_F} dr'\right](i+1)$$

$$g(r) = J_1(\tfrac{E_F r}{v_F}) \exp\left[-\int^r \tfrac{\Delta(r')}{v_F} dr'\right](i-1) \quad \text{(Eq.S4)}$$

We use a step function of $\Delta(r)$ ($\Delta = 1.8$ meV ($r > 8$ nm); $\Delta = 0$ meV ($r < 8$ nm)) in the equation and derive the $|u|^2$

$$|u|^2 = C \sum_{i=0}^{1}\left[J_i(\tfrac{E_F r}{\Delta_0 \xi_0}) e^{-\tfrac{(r-80)}{\xi_0}}\right]^2 \quad \text{(Eq.S5)}$$

where $J_i(x)$ is the Bessel function, $\Delta(r)$ is the superconducting paring potential, $E_F$ is the Fermi energy, $v_F$ is the Fermi velocity, $\Delta_0$ and $\xi_0$ are the superconducting gap and the coherence length at zero temperature, respectively. At temperatures higher than the level broadening of Majorana, the tunneling current is proportional to $|u|^2$.

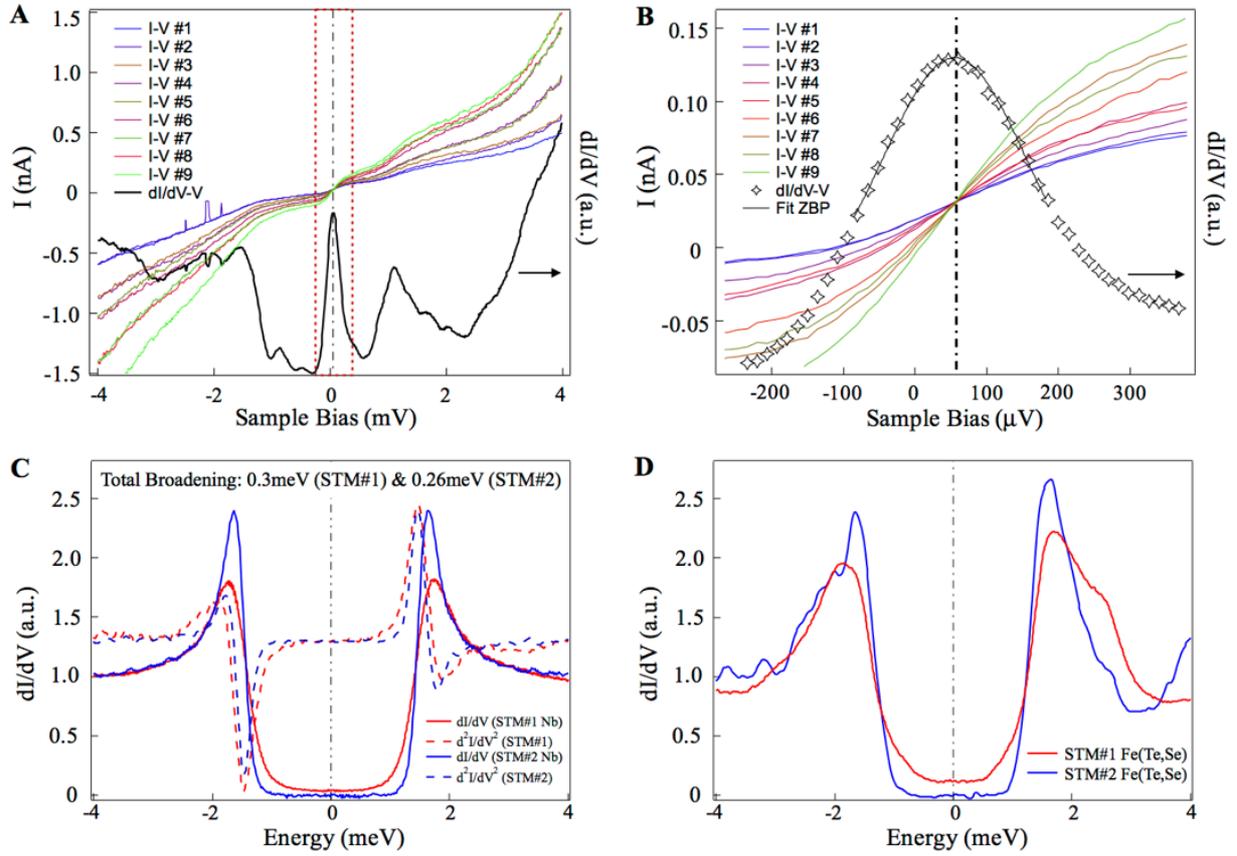

Fig. S1. **Voltage offset and energy resolution calibrations.** **(A)-(B)** Raw STS data before voltage offset calibration. **(A)** Simultaneous spectra measured at a vortex core center. There are 9 *I-V* curves with different values of tunneling resistance, the overlapping point which has a true zero current should have a true zero voltage. The *dI/dV* curves, obtained from lock-in techniques simultaneously, showing its peak position exactly at this overlapping point. **(B)** A zoom-in plot of the red dash box in A, showing that this particular voltage offset is 57.7 µeV. The energy step of the *dI/dV* curve is 14 µeV. **(C)-(D)** Superconducting spectra after their voltage offset calibrations measured by two LT-STM systems, showing the particle-hole symmetry of the superconducting gap. **(C)** Superconducting gap of a Nb single crystal measured at 0.4 K / 0 T. The total broadening of the superconducting gap edge is 0.3 meV for STM#1 and 0.26 meV for STM#2. **(D)** Superconducting gap of FeTe$_{0.55}$Se$_{0.45}$ single crystals measured at 0.4 K / 0 T which can form a MBS inside a vortex core under the magnetic field.

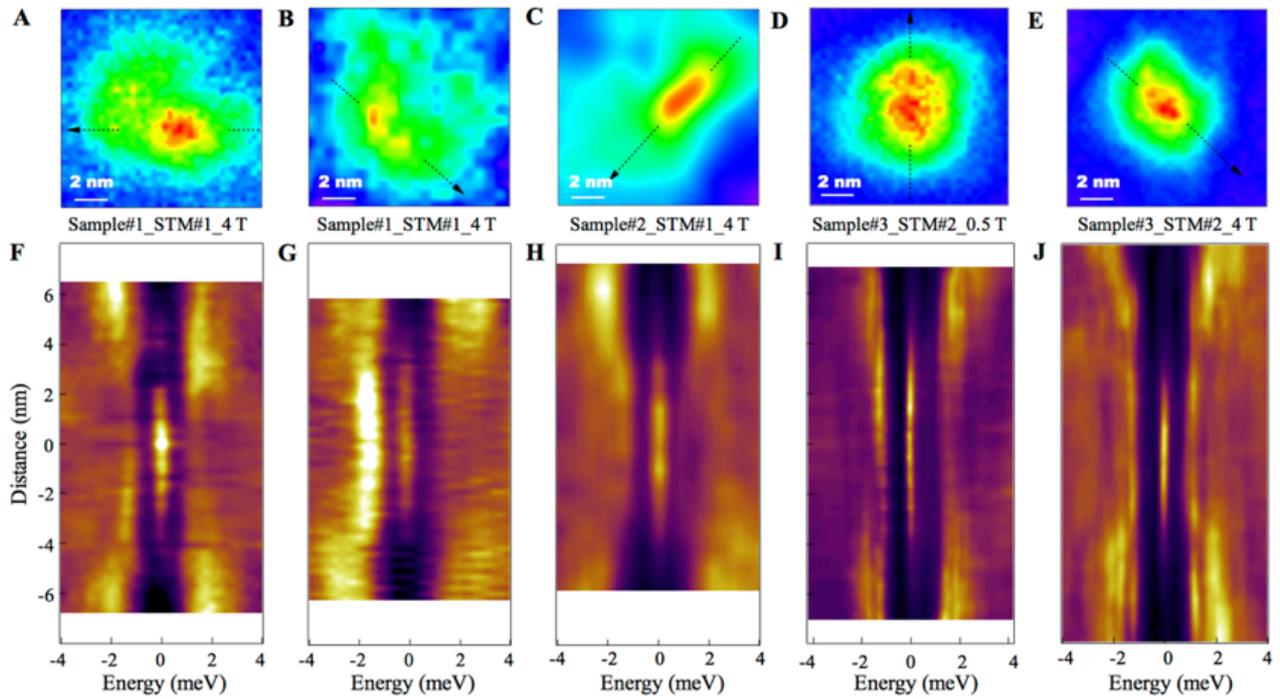

Fig. S2 **Spatial line profile reproducibility under different samples, magnetic fields, and equipment.** **(A)-(E)** ZBC maps (area = 12 nm × 12 nm) around vortex cores. **(F)-(J)** line-cut intensity plots along the black dash line indicated in (A)-(E), respectively. The data are normalized by integrated area of each *dI/dV* spectra.

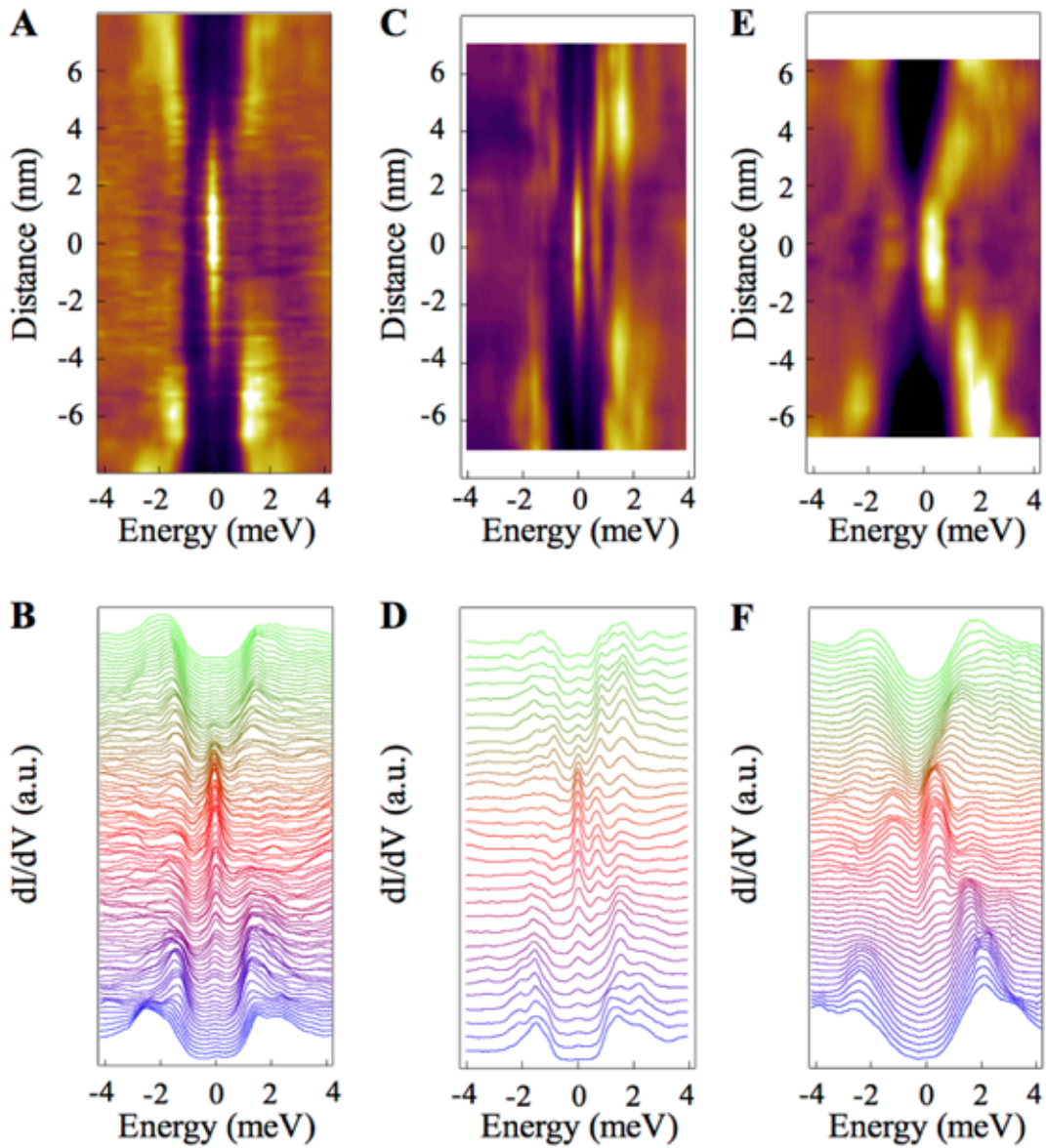

Fig. S3 **Three types of bound states inside vortex cores of FeTe$_{0.55}$Se$_{0.45}$.** **(A)-(B)** Linecut and waterfall plots of pure Majorana bound states, which are reploted from the middle panel of Figs. 3F and 3H (4 T). **(C)-(D)** Linecut and waterfall plots of co-existence cases. Both non-dispersive MBS peaks and dispersive CBS peaks (~ 0.7 meV) were observed (6 T). **(E)-(F)** Linecut and waterfall plots of CBS. In this case no zero-bias peaks were observed (4 T). All the data were measured at 0.55 K.

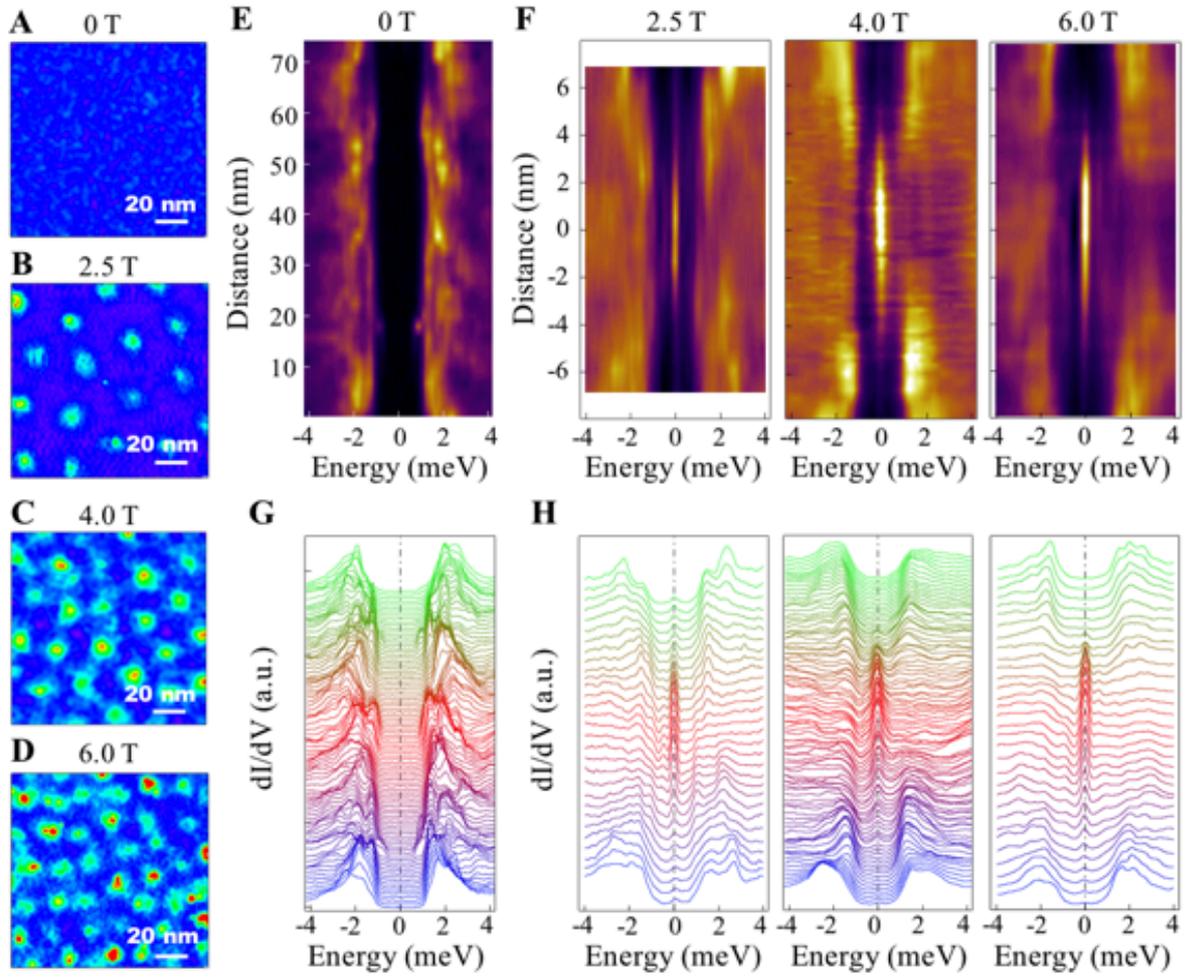

Fig. S4 **Majorana bound states under different magnetic fields.** (A)-(D) 120 nm × 120 nm ZBC maps at different magnetic fields. Inter-vortex length decreases with increasing magnetic field, that is, 70 nm for 0.5 T (Fig. 1E), 32 nm for 2.5 T, 25 nm for 4.0 T, and 20 nm for 6.0 T. They are consistent with theoretical prediction on Abrikosov vortex lattice, that is, 68 nm for 0.5 T, 30 nm for 2.5 T, 24 nm for 4.0 T, and 19.6 nm for 6.0 T. (E)-(F) Line-cut intensity plots at 0 T, 2.5 T, 4.0 T, and 6.0 T respectively. (G)-(H) Corresponding waterfall plot at 0 T, 2.5 T, 4.0 T, and 6.0 T. Zero magnetic field spectra show full superconducting gap and spatial inhomogeneity of $FeTe_{0.55}Se_{0.45}$. Note that a weak CBS coexists with the MBS at data of 2.5 T (6.0 T). The CBS peak locates at about - 0.5meV (+ 0.6 meV) in the core center and disperses toward the gap edge away from the core center. Settings: 4.0 T data measured by STM#1 with settings, $V_s = -10$ mV, $I_t = 100$ pA, and $T = 0.55$ K, while other data is measured by STM#2 with $V_s = -5$ mV, $I_t = 200$ pA.

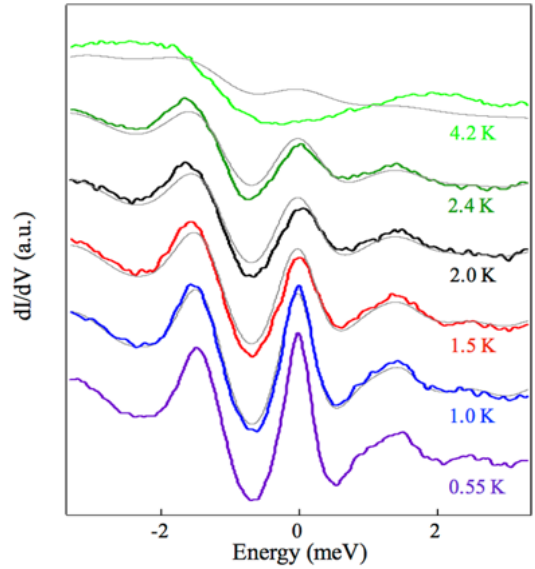

Fig. S5 **Absence of zero bias peak at 4.2 K which is not a simply result of thermal broadening.** Data under different temperatures (replot of Fig. 3D in the main text). The gray curves are numerically broadened 0.55-K spectra at each temperature. A convolution between the 0.55-K spectrum and the related Fermi-Dirac function is used to check the effect of thermal smearing out, which shows there is an additional mechanism contribution to the suppression of the zero-bias peak at higher temperature.

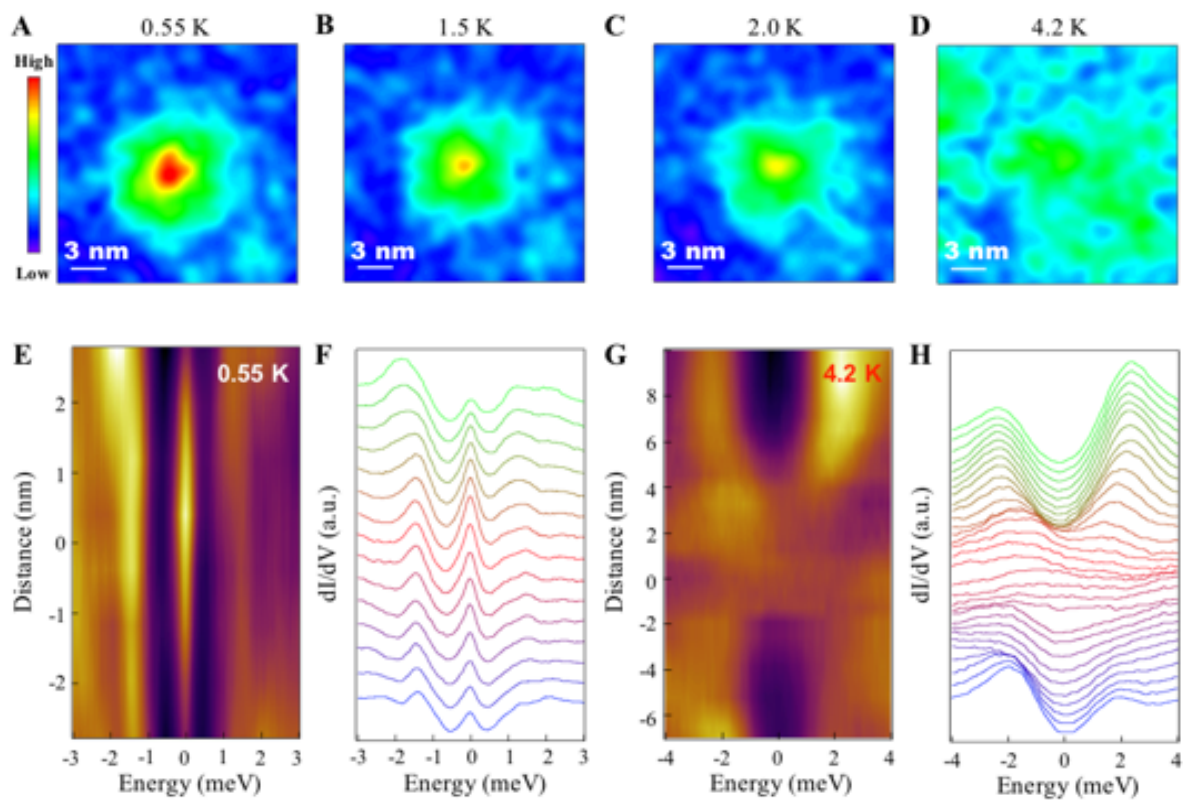

Fig. S6 **Line profile under low and high temperature.** **(A)-(D)** ZBC map at 0.55 K, 1.5 K, 2.0 K, and 4.2 K respectively. **(E)** Typical low temperature line profile at 0.55 K. **(F)** Waterfall plot of (E). **(G)** Typical higher temperature line profile at 4.2 K. **(H)** Waterfall plot of (G). All the data were measured at 4.0 T.

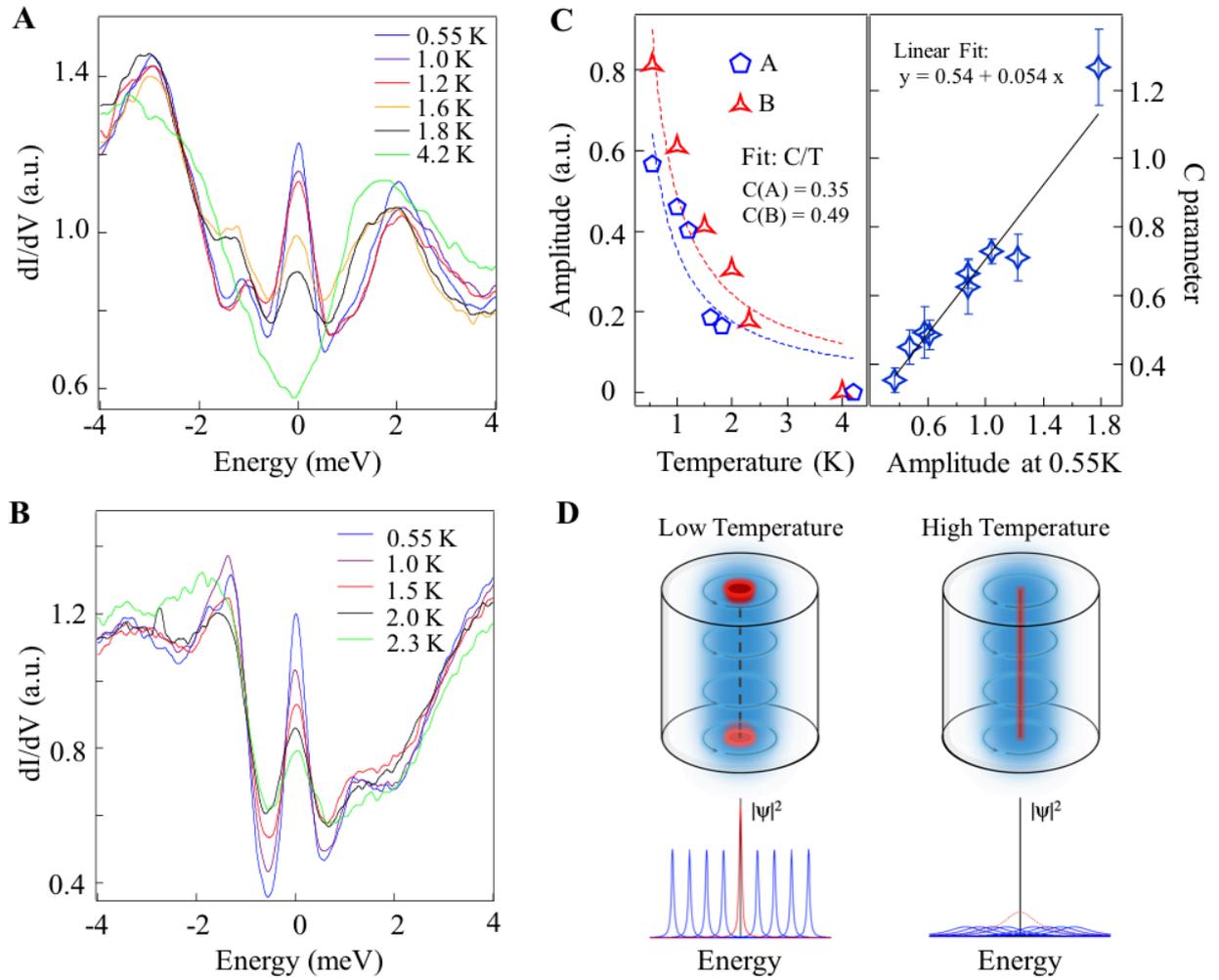

Fig. S7 **More examples of MBS temperature evolution.** **(A)-(B)** Two examples of Majorana ZBPs at the center of vortex core under different temperatures. **(C)** Left panel: C/$T$ fitting of amplitude of Majorana ZBPs under different temperatures. Red triangles are extracted from spectra in Fig. S7B. Blue hexagons are extracted from spectra in Fig. S7A. Right panel: summary on several temperature evolution measurements. The fitting parameter C, a characteristic temperature indicating MBS robustness against thermal influence, is positively correlated with the ZBP amplitude measured at 0.55 K. This indicates stronger Majorana ZBPs can survive at higher temperature. **(D)** Schematic explanation of the temperature effect on Majorana ZBPs. Upper row: schematic of a vortex with a vertical magnetic field applied at low and high temperature. Lower row: corresponding schematic of the energy-resolved bound states inside the vortex line (Red solid curves: Majorana ZBPs at the ends of the vortex line; blue curves: CBSs inside the bulk vortex line. All data sets are measured under 4 T and $I_t$ = 100 pA, $V_s$ = -10 mV. (Tunneling barrier conductance: $G_N \equiv I_t / V_s = 1.29 \times 10^{-4} \times 2e^2/h$)

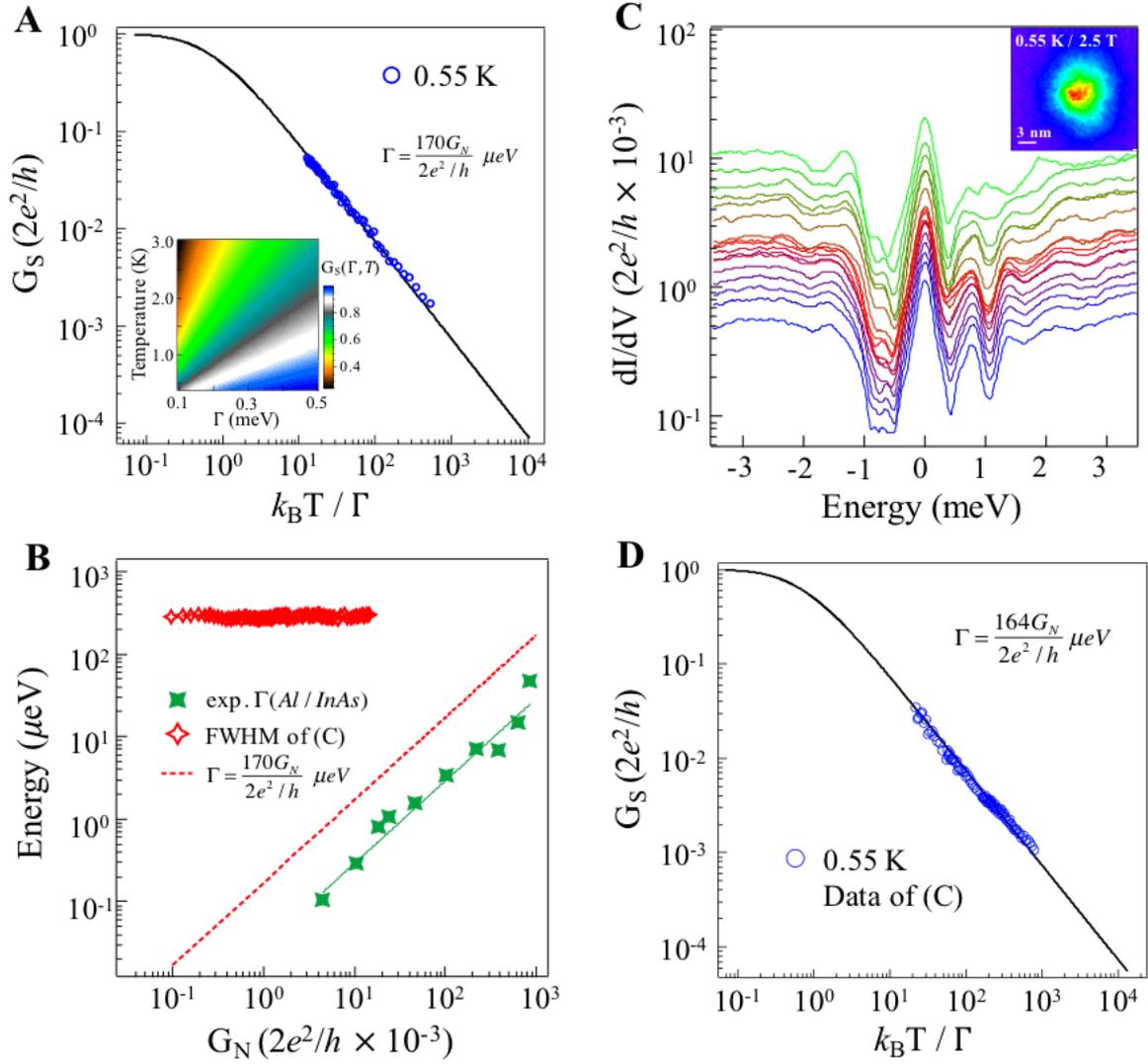

Fig. S8 **Additional Results of tunneling barrier evolution.** (A) Scaling behavior of the zero-bias conductance (ZBC) shown in Fig. 3A of the main text. Black line: Calculated ZBC of a single Majorana mode at low temperature using Eq. S2; Blue circles: ZBC of spectra at 0.55 K. Inset figure: calculation plot of $G_s(\Gamma, T)$. (B) Red dash line is a plot of tunneling broadening following the formula inset. Red stars are FWHM of spectra in (C). The average value of FWHM is about 0.28 meV. Green stars are tunneling broadening in nanowire devices. It is extracted from Fig. 3A in Ref. 14. (C) Another set of data about tunneling barrier evolution of MBS measured at 0.55 K and 2.5 T. MBSs locate at zero energy over two orders of magnitude in tunneling barrier conductance. Inset figure: ZBC map. (D) Scaling behavior of the ZBC shown in Fig. S8C.

**Table S1. Band Parameters of Fe(Te, Se)**

| Orbital/Band | $\Delta$ (meV) | $E_F$ (meV) | $\Delta^2/E_F$ (meV) |
|:---:|:---:|:---:|:---:|
| $\Gamma$ (TSS) | 1.8 | 4.4 | 0.74 |
| $\Gamma$ | 2.5 | 5~30 | 0.21~1.25 |
| M | 4.2 | 15~40 | 0.44~1.18 |